\documentstyle[epsf,11pt]{article}

\begin{document}

\title{Multiwalled Carbon Nanotubes as Building Blocks in Nanoelectronics}
\author{Markus Ahlskog, Pertti Hakonen, Mikko Paalanen, \\Leif Roschier, and Reeta
Tarkiainen \\
\\Low Temperature Laboratory, \\Helsinki University of Technology, \\FIN-02015
HUT, Finland}
\date{\today}
\maketitle

\begin{abstract}
Molecular level components, like carbon
multiwalled nanotubes (MWNT), show great potential for future 
nanoelectronics. 
At low frequencies, only the outermost carbon layer determines
the transport properties of the MWNT.  
Due to the multiwalled structure and large 
capacitive interlayer coupling, also the inner layers 
contribute to the conduction at high frequencies. 
Consequently, the conduction properties of MWNTs 
are not very far from those of 
regular conductors with well-defined electrical characteristics. 
In our work we have experimentally utilized this 
fact in constructing various nanoelectronic 
components out of MWNTs, such as single electron 
transistors (SET), lumped resistors, and transmission lines. 
We present results on several nano- tube samples, 
grown both using chemical vapor deposition as 
well as arc-discharge vaporization.
Our results show that SET-electrometers with a noise 
level as low as $6\cdot 10^{-6}$ $e/\sqrt{\textrm{Hz}}$ 
 (at $45$ Hz) can be built using arc-discharge-grown 
carbon nanotubes. Moreover, short
 nanotubes with small contact areas are found to work 
at $4.2$ K with good gate modulation. Reactive ion 
etching on CVD tubes is employed to produce nearly 
Ohmic components with a resistance of $200$ k$\Omega$ over a 
$2$ $\mu$m section. At high frequencies, MWNTs work over 
micron distances as special LC-transmission lines
 with high impedance, on the order of $5$ k$\Omega$.

\end{abstract}


\section{INTRODUCTION}

The large number of organic compounds and living organisms is based on the
unique chemical properties of elemental carbon. Carbon forms a variety of
compounds and, in pure form, exists in diamond as well as in graphite
structures. Despite these well known properties of carbon, the discovery of
fullerenes and nanotubes took the science community by surprise. The carbon
nanotubes were discovered by Iijima in 1991 in Japan \cite{Iijima}. It was
soon realized that both semiconducting and metallic nanotubes should exist 
\cite{Hamada}. The mass production of nanotubes succeeded in 1992 \cite
{masspro} and the first electrical transport measurements in 1996 
\cite{Langer96}. Various possible applications of nanotubes have quickly
emerged, for example in molecular electronics.\cite{World} The Coulomb blockade was
detected in transport measurements in 1997,\cite{Tans,bock}
demonstrating that the nanotubes are suitable building blocks of single
electron transistors (SETs).

The single electron transistor (SET) has, as an ordinary transistor, three
terminals: the source, the drain and the gate. In addition, the SET contains
a nearly isolated island between the source and the drain contacts. The
principle of the SET is based on the repulsive interaction between electrons
which becomes important when the island is smaller than about one micron and
its contacts to source and drain are weak (resistance larger than the
quantum of resistance $R_{Q}=h/e^2 \approx $ 26 k$\Omega $). 
The first successful SET was made by Fulton and
Dolan\cite{Fulton} at AT\&T Bell Laboratories in 1987 using aluminum thin film
technology and electron beam
lithography. Since the early success
with the Aluminum-SET, SETs have also been manufactured out of
two-dimensional electron gas of SiMOSFETs, with metal nanoparticles
and lately with carbon nanotubes. In
case of carbon nanotubes, the nanotube forms the central island and
the relatively poor contact resistance between the leads and the nanotube
ensures the required isolation.

In this paper we present work on basic electronic devices made of MWNTs.
As a building block of a nanoscale electronic device, the carbon nanotube
may fill many functions. Here we introduce different cases, based on our own
measurements, where a multiwalled carbon nanotube is used either 
as the island of a SET, 
as an Ohmic resistor, or as a transmission line. Furthermore, 
we compare devices made
of nanotubes that were synthesized using different methods. 

\bigskip 

\section{BASICS OF NANOTUBES}

\subsection{Electronic properties}

Graphite is made out of fairly loosely connected two dimensional carbon
layers, each layer having hexagonal lattice structure. Carbon
nanotubes are formed from such graphitic layers.\cite{book} 
A single walled nanotube (SWNT)
is formed when a piece of graphite sheet is wrapped into a cylindrical form,
the edges are seamlessly joined together and the ends of the cylinder
are closed. A multiwalled nanotube (MWNT) is made out of several concentric
graphite cylinders. The SWNTs are about 1-2 nm in diameter. The MWNTs, in
contrast, can be as large as 20 - 30 nm in diameter and are more convenient
for experiments. In constructing a SWNT one has several possibilities to
cut the original graphite sheet: The width and the length of the sheet can
be varied as well as the angle between the symmetry axis of the hexagonal
sheet and the main axis of the final cylinder. The electrical properties of
the nanotube are sensitive to the orientation of the hexagonal lattice because
it determines the density of electron states at the Fermi level.

The band structure of graphite was calculated already in 1947 by Wallace
using the tight binding approximation\cite{Wallace}. 
A graphite sheet is a semimetal,
whose two-dimensional band structure near the Fermi surface consists of
six conical energy surfaces in the first Brillouin zone. Fermi surface of
the undoped graphite consists of six points, the vertices 
of the six cones. When
the two dimensional graphite sheet is wrapped into cylindrical form, the
transverse electron motion around the cylinder surface becomes quantized. The
corresponding transverse energy levels of SWNTs are separated by about 1 eV
and only the lowest band is occupied at room temperature and below it. The
electrons on the lowest band move along the cylinder axis and behave truly
one-dimensionally. One can make nanotubes which are several microns long.
The longitudinal quantization of the electron motion leads to the fine
structure of the energy bands on the energy scale of 1 meV. This fine
structure is, however, washed away by the broadening of the levels due to
impurity scattering.

Calculations show that undoped SWNTs are either semiconductors or
metals, approximately one third of the tubes having the ''metallic''
orientation between the underlying hexagonal structure and the tube axis. In
the semiconducting nanotubes there are no electron states at the Fermi level
and the band gap is several hundreds of meVs. These tubes are good
insulators at small bias voltages. In the metallic nanotubes one-dimensional
energy bands cross the Fermi level. They are constructed out of the six
energy cones of the graphite. Because three of the cones are equivalent, the
six energy cones of graphite collapse into two independent one-dimensional
conduction channels. Taking into account the spin degeneracy of the
electrons, the metallic SWNTs have altogether four independent conduction
channels. The number of conduction channels is important because it
determines the electrical conductivity of the nanotube. The conductivity of
a ballistic one-dimensional system is given by $e^{2}M/h$,
where $e$ is the charge of electron, $h$ Planck's constant and $M$ the
number of independent conduction channels. Thus the resistance of a
ballistic nanotube is expected to be 6.45 k$\Omega $ ($M=4$).

The concentric layers of a MWNT are estimated to be in poor electrical
contact to each other. Some experimental results have been interpreted with the
assumption that only the outermost layer participates in the electron
conduction.\cite{Schonenberger99} Therefore, 
one expects the electrical properties of the MWNTs
and SWNTs to be similar. However, the metallic inner layers can contribute
to the electrical properties of the MWNT at high frequencies.\cite{Sonin}
In fact, the analysis of tunneling experiments in MWNTs has recently turned out to be
more difficult than expected.\cite{TarkiainenPRL,Bachtold00,Egger00}

\subsection{Synthesis of multiwalled carbon nanotubes and their properties}

Carbon nanotubes are synthesized with various methods that differ from each
other in crucial ways with regard to the growth conditions, resulting in
nanotubes of different quality. In MWNTs in particular, various defects can
appear that distort the structure of the ideal MWNT. 
In the arc-discharge\cite{masspro}\ (AD) and 
related methods nanotubes are produced in an
inert-gas atmosphere from graphite at such a high local temperature that
the carbon evaporates and subsequently forms the nanotubes. Later,
catalytical synthesis of nanotubes was started using chemical vapor
deposition (CVD) methods \cite{CVD}. In the CVD technique the nanotubes grow
from a catalyst over which a carbon-containing gas is led. The upper
temperature is limited by the requirement that the gas is not decomposed by
itself. Therefore the growth temperature with the CVD technique is lower than
with the other techniques. The lower growth 
temperature of CVD tubes is considered
as the main cause for their more defective structure.

\begin{figure}[tbp]
\begin{center}
\epsfxsize=120mm \hspace{0mm}\epsffile[20 20 590 200]{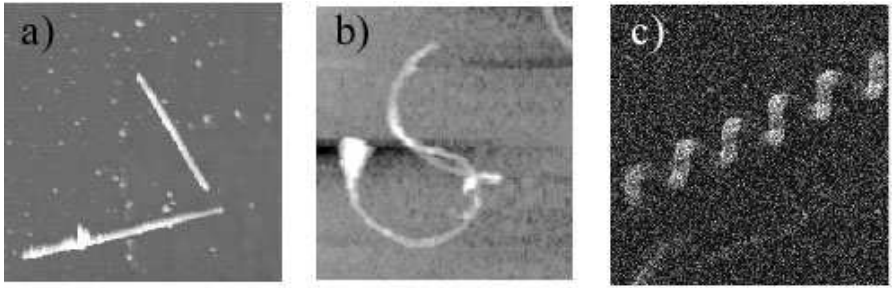}
\end{center}
\caption{AFM images of typical a) arc-discharge and b) CVD-grown nanotubes.
c) SEM image of a CVD-grown helical nanotube (Courtesy of E. Seynaeve,
VSM, Kath. Univ. Leuven, Belgium).}
\label{SEM}
\end{figure}

AD tubes are generally straight, as shown in Fig.~\ref{SEM}a, and exhibit
a rather flawless structure in TEM images. Indeed, these are rather good
approximations of the ideal nanotubes whose basic electronic properties were
discussed above. Multiwalled CVD tubes, on the other hand, often exhibit a
significant amount of disorder in TEM analysis, although images of well
ordered sections of individual CVD tubes have also been 
reported \cite{CVDimage}. It is likely
that CVD material exhibits large variations in quality. 
This variation is evident in our experiments on CVD tubes treated with
reactive ion etching (RIE).
In a short etching time most tubes were uniformly thinned
by a factor of 2, while some tubes had vanished almost completely. The
defects in a low-quality tube are much more susceptible to
an etchant than the regular graphitic walls. As a result of the defective
structure, CVD tubes are generally curved when observed either as
freestanding in a TEM or as deposited on substrates, as shown in the AFM
image of Fig. \ref{SEM}b. This curvature is rigid and often extends into
three dimensions. The nanotube (diameter 20 nm) in  Fig.~\ref{SEM}b has a
tail to the left
which extends more than 100 nm above the substrate, despite the significant
van der Waals forces that exist between the substrate and the
nanotube. Note that the tail appears broadened due to the interaction
between the AFM tip and the upwards pointing tail section. 
The most impressive of the three dimensionally extended tubes are
the regularly coiled 3D spirals of multiwalled CVD-grown tubes \cite{Spiral}, as
shown in Fig.~\ref{SEM}c. Salvetat {\it et al.}\cite{Salvetat} 
presented work
comparing the elastic moduli of AD and CVD multiwalled tubes, concluding
that the former are significantly stiffer than the latter. Thus, although AD
tubes have generally better electronic performance from the point of view of
device physics, CVD tubes have their own interesting and
useful properties.

\section{CONSTRUCTION OF NANOTUBE CIRCUITS}

For electronic transport measurements individual nanotubes have to be
connected to external leads. This is not a trivial task because the
nanotubes are quite small, difficult to image and hard to move around. In
making simple nanotube circuits one starts from a solution (actually a
dispersion). A droplet of this
solution is placed on the substrate at a desired location, and the solvent is
evaporated away. This leaves behind a number of randomly placed nanotubes on
the surface of the substrate. There are two possibilities to connect the
nanotubes to the leads. In the first method the leads are evaporated on top
of the nanotubes. This method requires accurate imaging of the nanotubes and
sophisticated alignment techniques when making the evaporation mask. In the
second method one moves the nanotubes on top of prefabricated gold
electrodes (or makes a large array of electrodes in the hope of finding one
nanotube already extending over at least two electrodes). Both of these
methods are hampered by poor electrical contacts between the nanotubes and
electrodes. This is a
problem especially in the case, where nanotubes are deposited on top of the
electrodes. The contact resistances range from a few tens of k$\Omega $ to
gigaohms and are not well understood. There are some methods to improve the contacts. 
Bachtold {\it et al.} \cite{bach} successfully used electron bombardment in SEM, 
while Jeong-O Lee {\it et al.}\cite{namkim} used quick heating at high temperature. 
We have also used this method in some cases.
Contact resistances decrease considerably by a 30 second annealing 
of the sample in vacuum at 700$^{\circ }$C. We also have verified the 
mechanical change in a contact caused by heating: to move the tube after heating 
required considerably larger applied force. 

Usually an atomic force microscope (AFM) is used for imaging surface topography
with a tip that has a radius of curvature down
to 10 nm.  However, it can also be used to move small objects along the substrates. 
We have developed an
AFM based manipulation method where both moving and imaging are done 
in the non-contact mode.\cite{Martin}
In this mode one probes the attractive van der Waals force
between the vibrating AFM tip and the surface. Using our method, the location
of the object can be seen during the movement by monitoring the cantilever
oscillation amplitude. This is how we can build
electrical circuits containing several nanotubes, or move nanotubes on top
of gold electrodes. 

\begin{figure}[tbp]
\begin{center}
\epsfxsize=100mm \hspace{0mm}\epsffile[20 20 583 180]{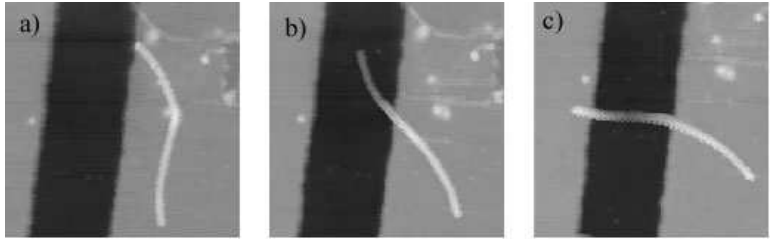}
\end{center}
\caption{A sequence of AFM-images illustrating smooth bending  
of a CVD-synthesized tube (length 1.3 $\mu$m) during manipulation.
According to Figs. 2b and 2c, the tube is in contact with the SiO$_2$
substrate during the translation process across the gap between the
25 nm thick gold electrodes.}
\label{Manipulation}
\end{figure}

The differences between the arc-discharge and CVD tubes become clearly visible 
in our AFM manipulations of these tubes. When an AD tube is pushed with an AFM tip,
it usually
either moves as a whole, without bending, or bends sharply at a
certain point, with
the rest of the tube staying on its place. These
conclusions are similar to those of Falvo {\it et al.} \cite{Superfine}, who
studied manipulation of AD tubes more extensively. We have found that
manipulation of those CVD tubes that have significant 3D bending out of the 
surface is very difficult due to the sticking of the tubes to the
AFM tip. Also the van der Waals forces that hold the nanotube in place and stabilize
strained configurations are reduced. The manipulation of the CVD tubes is also
different in the sense that the bending is continuous, without any pivot points, as
shown in the manipulation sequence of a CVD tube in Fig. \ref{Manipulation}.
This is of course expected due to the smaller stiffness of CVD tubes. Thus, 
only manipulation of AD tubes is convenient, when the required distance is 
several micrometers.

Bending of a nanotube will cause major local changes in the 
electronic structure, and consequently in intratube transport, 
as has been shown in calculational work.\cite{Marco} These 
references, however, deal basically with single shell nanotubes. On the 
experimental side, a large increase in resistance has been 
observed for SWNTs upon bending \cite{SWNTbend}. For MWNTs, 
the effect of bending is quite small compared to the total 
conductance of the tube, as found by Paulson {\it et al}.\cite{Paulson} 

\begin{figure}[tbp]
\begin{center}
\epsfxsize=100mm \hspace{0mm}\epsffile[20 20 590 190]{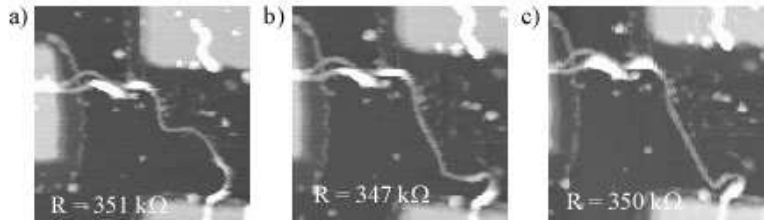}
\end{center}
\caption{The change in shape of a 1.7 $\mu$m long 
CVD nanotube has only a small
 effect on the conductance. The measured resistance is marked in each
 frame.}
\label{Figres}
\end{figure}

To observe the effect of bending to the resistance of a nanotube, 
resistance measurements can be done {\it in-situ} under the AFM.
We have, for example, studied one sample, where a MWNT (of 
length 1.7 $\mu$m) is below the electrodes (Fig. \ref{Figres}). 
The tube is CVD synthesized, as can be seen from its shape. 
During a long sequence of small movements, we observed no 
dramatic changes: the resistance varied between 340 and 
380 k$\Omega$, the initial value. Both increase 
and decrease of resistance, typically a few k$\Omega$, was observed 
between 
adjacent configurations. Based on our experience 
on CVD tubes, we expect that a sizable share of the resistance 
is in the tube itself. Thus we conclude that modest bending of 
a CVD grown MWNT has only a small effect on the intratube 
resistance. The result becomes more general when we note 
that similar conclusions were obtained\cite{Paulson} for 
AD grown MWNTs. It is not surprising that it is more 
difficult to bend a multiwalled than a single walled 
tube so sharply that a tunnel junction would emerge at the site of bending.

\begin{table}[ht]
\caption{Basic characteristics of our samples. The labeling indicates
 the nanotube synthesis method. The length $L$
 refers to the length between the electrodes. Zero-bias resistances are given
 for two point configuration at $T=300$ K and 4.2 K. The small ratio
 of $R$(4.2 K)/$R$(300 K)$= 1-7$ indicates that our samples are
 metallic except for AD2 which is semiconducting.}
\label{tab:table}
\begin{center}
\vskip 4mm 
\begin{tabular}[l]{|c|c|c|c|c|}
\hline
\hbox to 2 cm{\hfil Sample\hfil} & \hbox to 2 cm{\hfil Comment \hfil} & %
\hbox to 2 cm{\hfil $R$(300 K) \hfil} & \hbox to 
2 cm{\hfil $R$(4.2 K) \hfil} & \hbox to 2 cm{\hfil $L$ ($\mu$m)\hfil} \\ 
\hline\hline
CVD1 &  & 230 k$\Omega$ & 1.6 M$\Omega$ & 5.3 \\ \hline
CVD2 & RIE & 23 k$\Omega$ & 31 k$\Omega$ & 0.3 \\ \hline
CVD3 & RIE & 149 k$\Omega$ & 190 k$\Omega$ & 1.5 \\ \hline
CVD4 &  & 50 k$\Omega$ & 91 k$\Omega$ & 0.5 \\ \hline
AD1 & Cross & 71 k$\Omega$ & 170 k$\Omega$ & 1.7 \\ \hline
AD2 & & 5 M$\Omega$ & $\sim$ 200 M$\Omega$ & 0.3 \\ \hline
AD3 & & 28 k$\Omega$ & 84 k$\Omega$ & 0.3 \\ \hline
\end{tabular}
\end{center}
\end{table}

\section{EXAMPLES OF SETs MADE OF MWNT}

\begin{figure}[tbp]
\begin{center}
\epsfxsize=120mm \hspace{0mm}\epsffile[20 20 590 180]{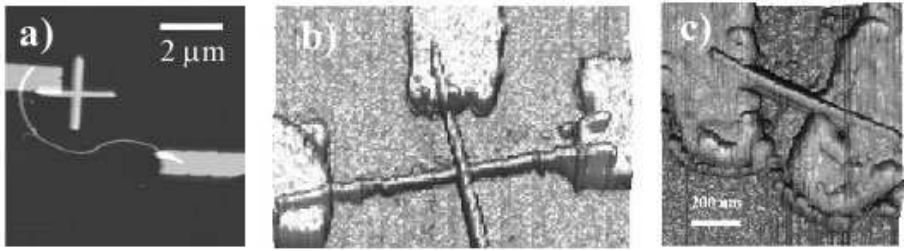}
\end{center}
\caption{ AFM images of some of our samples: a) CVD1 (Image size
  7$\times$7 $\mu$m$^2$), b) AD1 with two crossing
  multiwalled tubes ($\sim$ 1.3$\times$2.4 $\mu$m$^2$), and c) AD3 with a 
  free-standing section of 0.3 $\mu$m. Gate leads (not visible in
  the figures) of the same width as the other electrodes end at a
  distance of $1-5$ $\mu$m from the nanotube island.}
\label{Samples}
\end{figure}

We have fabricated MWNT-SETs both from CVD and AD tubes, some of which are
shown in Figure \ref{Samples}. Main characteristics of the samples are given in
Table 1. The nanotubes had typical diameters of 20 nm for the CVD tubes and
15 nm for the AD tubes. The length of the tube between the electrodes was
0.3-5 $\mu $m for the CVD and 0.3-1.7 $\mu $m for the AD tubes. Two of the
CVD tubes were etched in a RIE process for 15 seconds which reduced their
diameter uniformly from 20 nm to 10 nm. The electrodes were either
fabricated on top of the tubes (CVD1-CVD3 and AD1) or the tubes were
placed on top of
prefabricated electrodes (CVD4, AD2 and AD3). The electrodes were always made of
gold with a chromium sticking layer.

The quality of results obtained on our devices varied over a broad range. The
results on CVD tubes
were, in general, more irregular than those on AD tubes. Some of the
AD tubes turned out to be excellent islands for SETs.
As we will show below, the geometrical structure and the construction
method have a clear influence on the results. 

\subsection{Standard CVD-tube device}

Figure \ref{Samples}(a) shows an AFM image of device CVD1. It is made of a
6.8 $\mu$m long tube which has 
25 nm thick
gold electrodes evaporated on top of it. 
At room temperature, 
the resistance of the tube was 200 k$\Omega $. Below 4 K
 a zero-bias gap due to Coulomb blockade became clearly visible.

\begin{figure}[tbp]
\begin{center}
\epsfxsize=80mm \hspace{0mm}\epsffile[20 20 590 445]{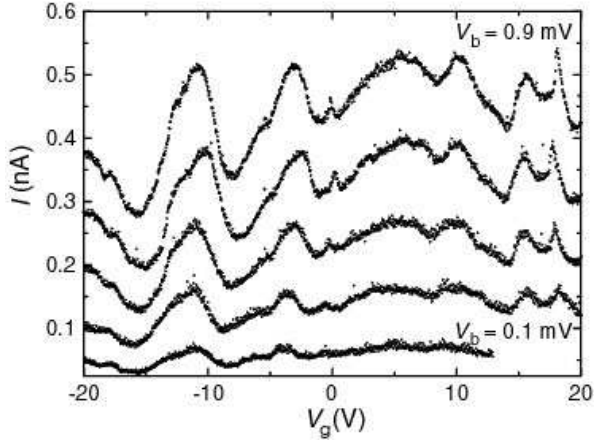}
\end{center}
\caption{Gate modulation curves (source-drain current {\it I vs.} gate
  voltage $V_g$) measured on the sample CVD1 at $T=0.8$ K. The bias
  voltage $V_b$ has been varied from 0.1 mV to 0.9 mV in steps of 0.2 mV. }
\label{CVDmod}
\end{figure}

A few gate modulation curves of this MWNT, measured at $T=0.8$ K, are 
shown in Fig. \ref{CVDmod}. Instead of the expected Coulomb blockade
period of 1V (estimated for the employed side-gate placed a few $\mu $m
away), only 
non-regular gate-modulation traces were found.
This irregular behavior points to the 
formation of a series of intra-tube quantum dots, caused by
disorder. 
 Disorder-induced splitting of a tube into several separate islands has been
suggested by McEuen {\it et al.}\cite{McEuen} to explain similar
experimental results in SWNTs.
From the data set of Fig. \ref{CVDmod} and other sets like it, 
we conclude that CVD tubes, at least with the present amount of disorder,
are not very suitable for central islands of SET electrometers.

\subsection{Arc-discharge tubes: Tube-gated device}

\begin{figure}[tbp]
\begin{center}
\epsfxsize=90mm \hspace{0mm}\epsffile[50 40 560 430]{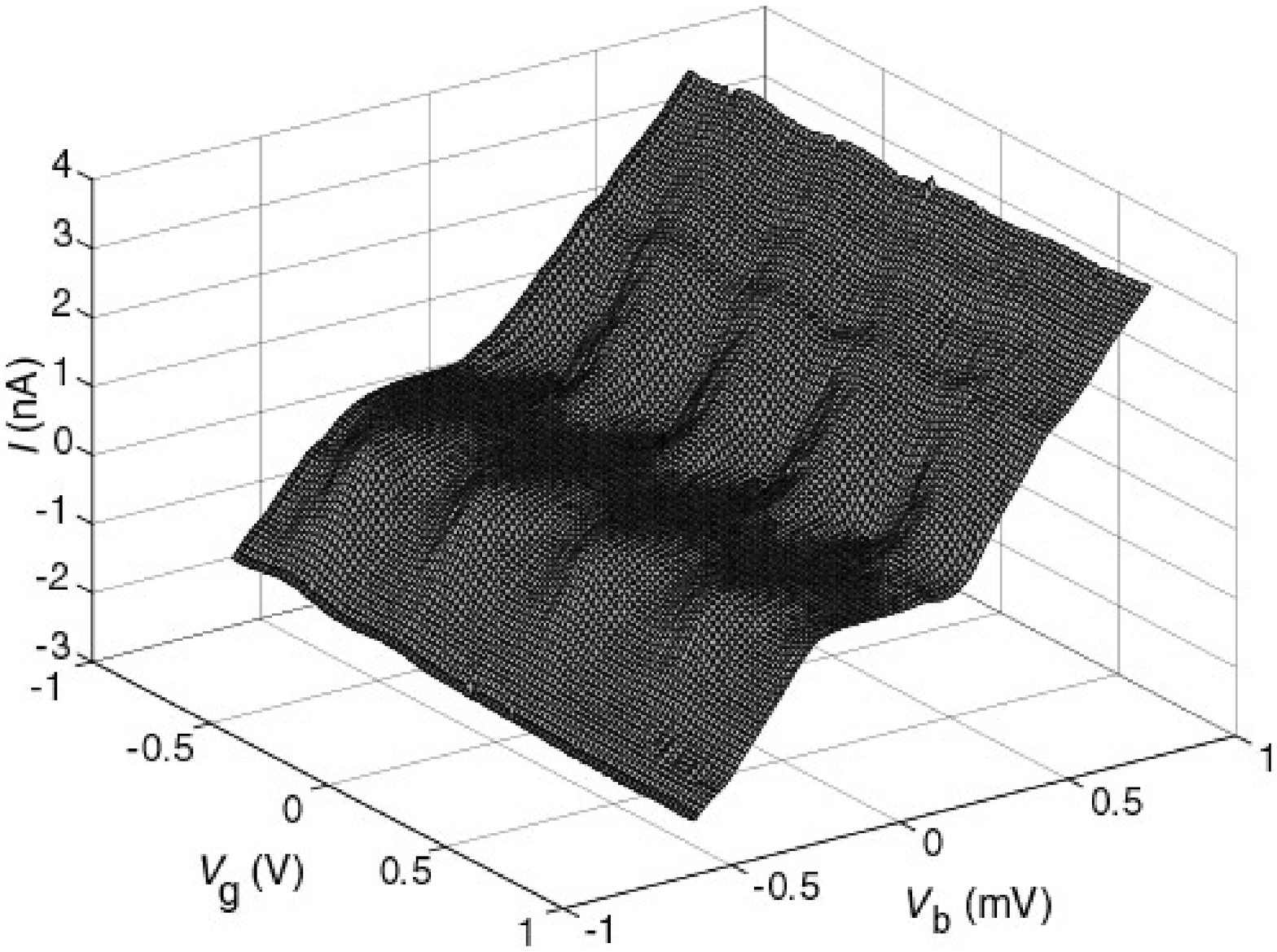}
\end{center}
\caption{Source-drain current $I$ of the lower tube in sample
  AD1 at $T=0.15$~K.\cite{Ahlskog} 
  The source-drain bias voltage is denoted by $V_b$ and the gate
  voltage by $V_{g}$. The SET blockade region is seen as the rhombic
  pattern in the center.}
\label{IVcross}
\end{figure}

In contrast to CVD-tube devices, SETs based on AD tubes display regular
$IV$-curve modulation with respect to the gate voltage.
Such regularity is particularly nicely illustrated by the results on 
the device of Fig. \ref{Samples}b
(AD1).\cite{Ahlskog} This sample is in fact composed of two crossing MWNTs: 
the lower tube, which acts as the central island of the SET, is
$2.3$ $\mu$m
long, and the crossing upper tube has been pushed into its position with AFM
manipulation. The two-point (zero-bias) resistance over the crossing was 
$\sim$ 10 M$\Omega $, which increased to $\sim1$ G$\Omega $ below 4
K. Thus, we
could utilize the upper tube for gating the current in the lower
tube in this construction. The device has a room
temperature resistance of 71 k$\Omega $. A Coulomb blockade develops fully
only at subkelvin temperatures, with a gap of about 1 mV at 150 mK. 
Figure \ref{IVcross} shows the source-drain current $I$ as a function of
the source-drain bias voltage $V_b$ and the gate voltage $V_{g}$, 
applied to a separate side gate.

In general, Fourier analysis of
the gate modulation curves revealed only one period, 
indicating the existence of only
one island. In contrast to CVD tubes, we conclude that this tube is not
broken into sections, neither by defects nor by the other MWNT placed on top of
it. 

The gate modulation period with the upper tube as a gate was measured as 
$\Delta V_{g}$ = 4 mV (using side gate $\Delta V_{g}$ = 440 mV). The
tube had a maximum modulation of 8 nA/$e$. We calculate
the gate capacitance to the upper tube as $C_{g}=e/\Delta V_{g}$ = 40
aF (for side gate $C_{g}$ = 0.4 aF). The present configuration with a crossing 
nanotube as a
large-capacitance gate electrode might be useful in certain SET applications for
reducing cross-talk between the gate and the other electrodes. Furthermore,
since the voltage gain of a current-biased SET is $C_{g}/C_{\Sigma }$ such a
construction allows for devices with high voltage gain
\cite{Visscher}. Here $C_{\Sigma}$ refers to the total island capacitance
including the tunnel junctions.

\subsection{AD tubes: 4.2-Kelvin device}

Short pieces of nanotubes,
equipped with small
contact areas, provide a straightforward way to reach SETs with large Coulomb energies.
Figure~\ref{Samples}c shows one of the shortest nanotube devices (AD3) that
we have made from AD tubes. In our short structures, 
the tubes have been manipulated
on top of gold electrodes (AD2 and AD3). The length of the tube between the
electrodes is 0.3 $\mu $m for both AD2 and AD3. Furthermore, the
latter is not touching the substrate between the electrodes. It is separated
from the underlying SiO$_{2}$ by 17 nm.
\begin{figure}[tbp]
\begin{center}
\epsfxsize=70mm \hspace{0mm}\epsffile[20 20 590 460]{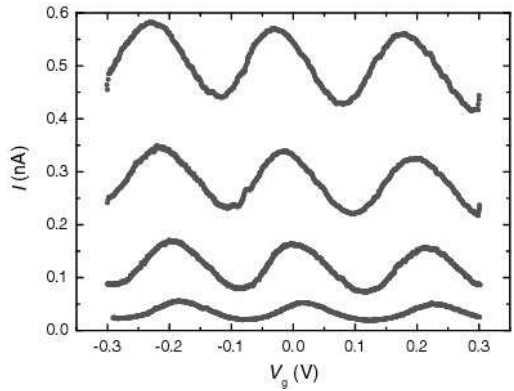}
\end{center}
\caption{Gate modulation curves measured on sample AD2 at
  $T=4.2$ K. Bias voltages are 8.0, 9.9, 11.6, 13.7 mV (from bottom to top).}
\label{GateMOD}
\end{figure}

The room temperature resistance of AD2 was rather high, 5 M$\Omega $, due to
weak gold-nanotube contacts, and increased to $\sim$ 200 M$\Omega $ at 4.2 K. Coulomb
blockade effects became clearly observable below a few Kelvin. Figure \ref
{GateMOD} shows gate modulation of the source-drain current at 4.2 K. The
modulation period is 200 mV, giving a gate capacitance of
0.8 aF. From measurements of the constant current curves we obtain the
junction capacitances and for the charging energy $E_{c}=e^{2}/2C_{\Sigma }$
= 2.1 meV. 
Thus this device is a rather simple implementation of a SET
working at the relatively high temperature of 4.2 K. Moreover, scaling down
the dimensions should be straightforward in order to minimize the island
size and, consequently, to raise the operating temperature even
further. 
The low temperature behavior of sample AD2 has been described
by Roschier {\it et al} \cite{RoschierAPL}.

\subsection{AD tubes: Low-noise device}

Nanotubes are quite susceptible to noise caused by charge trapping to
surface states.\cite{Zettle} Multiwalled tubes, fortunately, are not as
sensitive in this respect as single walled tubes  
even though some of our MWNT samples do show quite high noise
levels. We obtained the lowest
charge noise in device AD3 which had extraordinary properties in
several respects when compared with our other
samples\cite{Semiballistic}. Especially, the
transport in this tube was close to ballistic, which may be the
reason behind its good noise properties.

Sample AD3 had a room temperature 
resistance of 28 k$\Omega $. At subkelvin temperatures, we
measured a total resistance of $\approx $ 40 k$\Omega $ outside the Coulomb
blockade regime. The junction resistance of the nanotube-Au contacts are
thus less than the quantum of resistance $R_{Q}$ $\approx $ 26 k$\Omega $,
which means that the Coulomb blockade cannot fully develop. Consequently,
the Coulomb oscillations that we measure are smoothened.\cite{Semiballistic} 

As opposed to the usual Coulomb blockade,
nanotube AD3 exhibited increased conductance around zero bias, which we
attribute to resonant tunneling. Only two weakly quantized steps are seen, and
therefore this tube cannot be said to be fully ballistic. The ballisticity
of freestanding samples is likely to be enhanced most by
the decreased capacitive coupling between impurity states of the
substrate and the MWNT.

\begin{figure}[tbp]
\begin{center}
\epsfxsize=90mm \hspace{0mm}\epsffile[20 20 593 435]{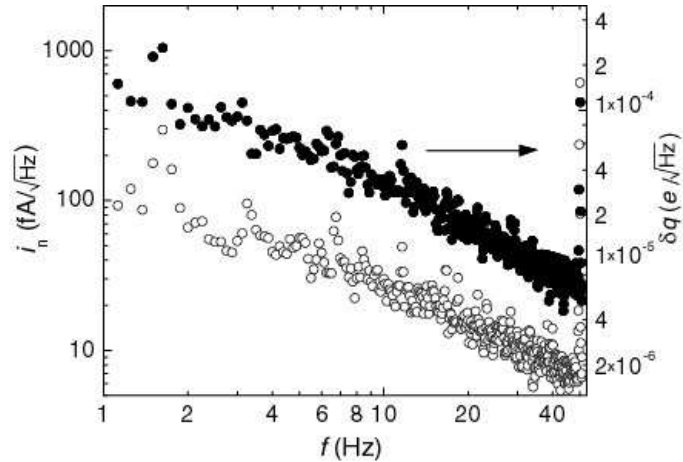}
\end{center}
\caption{Current noise $i_n$ measured on sample AD3 at $T=0.15$ K;\cite{Semiballistic}
  equivalent background charge fluctuation $\delta q$ is given on the right
  scale. Open circles denote the amplifier noise in the measurement setup. }
\label{Noise}
\end{figure}

Current noise $i_n$ of AD3, measured at a small voltage bias of 70 $\mu$V, 
is displayed in Fig. \ref{Noise}.
Frequency dependence of the noise power ($i_n^{2}$) has a $1/f^{2}$ character over the
range $5 < f < 50$ Hz. The
input equivalent charge noise $q_{n}$ is obtained from the measured current
noise according to the formula $q_{n}=C_{g}i_{n}/(\partial I/\partial
V_{g})$. In the Coulomb blockade regime a modulation of the noise was seen
as expected for a SET. At a frequency of 45 Hz, we obtain the charge noise $%
q_{n}=6\cdot 10^{-6}$ $e/\sqrt{\textrm{Hz}}$, which is comparable to the best
metallic SET devices reported to date \cite{Krupenin}. Theoretically the
minimum noise level for a SET is $q^{min}_n=\sqrt{\hbar C_{\Sigma }\Delta
fR_{Q}/R_{T}}$, where $\Delta f$ denotes the frequency
range and $R_{T}$ is the tunneling resistance \cite{Korotkov}. Taking $%
R_{Q}/R_{T}$ $\sim $ 1 and assuming no cotunneling, we obtain the minimum
noise as $1\cdot 10^{-6}e/\sqrt{\textrm{Hz}}$. This implies that white noise
would dominate over $1/f$ noise above 3 kHz. In fact, the shot noise limited
region has been reached in our most recent experiments.\cite{TarkiainenNoise}

\section{ MWNTs AS RESISTORS}

\begin{figure}[tbp]
\begin{center}
\epsfxsize=80mm \hspace{0mm}\epsffile[20 20 591 430]{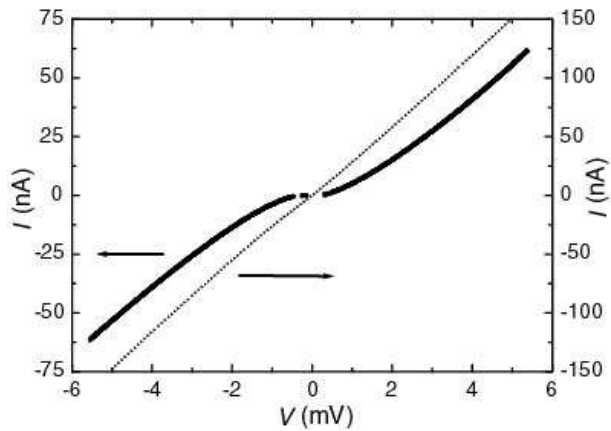}
\end{center}
\caption{$IV$-curves for samples CVD2 (right current scale) and CVD4
  (left current scale) illustrating the
  enhanced Ohmic behavior of CVD2 after the RIE treatment. Measurement
  temperature for both traces was $T=0.15$ K.}
\label{CVDrie}
\end{figure}

According to simple Drude-type arguments, disordered MWNTs are good
candidates for lumped resistive elements. As was already mentioned above, we
have tried to pursue this idea by selecting a couple of CVD tubes for 
RIE-etching in order to enhance the amount of disorder in them. 
The etching halved uniformly the diameter from 20 nm to 10 nm.
Gold electrodes were placed on top of these tubes. The
distance between them is 0.3 $\mu$m. Figure \ref{CVDrie} 
shows the IV
characteristics of one of these tubes (CVD2) at low temperatures,
compared with an untreated CVD-tube (CVD4) that has roughly the same
electrode spacing. At room temperature we measure a 2-point resistance
of 23 k$\Omega$ in the RIE-etched tube. At $T=100$ mK we
observe only a weak Coulomb blockade-type nonlinearity in the IV
characteristics and no discernible Coulomb oscillations, in contrast
with CVD4. We therefore assume
a relatively small contact resistance and consequently that the measured 2-point
resistance indicates the intrinsic resistivity of the RIE-etched MWNT to be
rather large, on the order of 100 k$\Omega$/$\mu$m. 

These results were confirmed in our second RIE-etched sample (CVD3), measured
in a 4-point configuration. With this sample we observed, in addition, 
that the resistance measured
between different electrode pairs scales roughly with the
distance, in
accord with Ohmic behavior. Presently we can not tell exactly what role the
RIE treatment had for the electronic properties of these tubes. Their
intrinsic resistance seems to be higher than that of the untreated CVD tubes by
a factor $\simeq 3$. 

The high intrinsic resistance together with Ohmic addition
rule for the nanotube resistance makes the RIE-etched CVD tubes simple nanoscale
Ohmic resistors. 
These resistors perform rather well compared with the standard
technology utilizing thin chromium films \cite{Penttila}. 
Typical thin Cr-wires can be
employed as linear resistances up to values 10 k$\Omega/\mu$m while RIE-etched tubes
yield 100 k$\Omega/\mu$m if their small nonlinearity in the zero-bias
region can be accepted.
With better metal-tube contacts, this non-linearity can presumably
be made even smaller.

\section{MWNTs AS TRANSMISSION LINES}

In the previous sections (Sects. 4 and 5)  we treated MWNTs
as lumped elements of simple electronic circuits. Howewer, their high aspect
ratio allows for nanotube lengths up to
macroscopic dimensions. Thus any longer section of a MWNT and nearby
metallic structures form a transmission line, where the nanotube acts as an
inner conductor. Because the electrons move nearly ballistically in a
metallic defect-free tube, one naively expects MWNTs to be excellent
transmission lines for high frequency electromagnetic waves. However,
rather the
opposite turns out to be true, since MWNTs form special, high impedance
transmission lines with a low propagation velocity \cite{TarkiainenPRL}.
 
\begin{figure}[tbp]
\begin{center}
\epsfxsize=80mm \hspace{0mm}\epsffile[20 20 581 465]{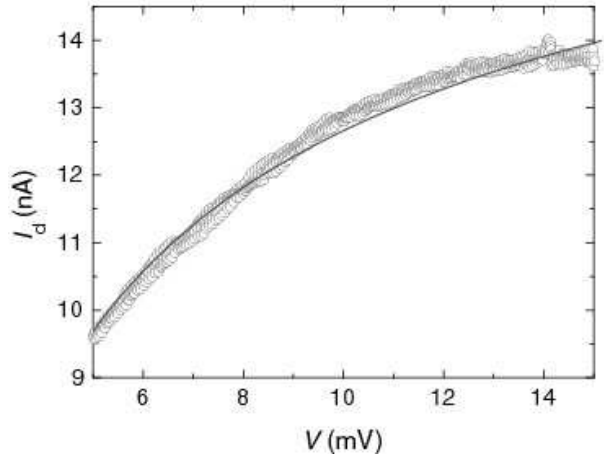}
\end{center}
\caption{Deficit current $I_{d}=V/R_T-I$ as a function of voltage
  $V$, illustrating the asymptotic approach towards Ohm's law
   in sample AD1 at $T=0.12$ K.}
\label{TailLIN}
\end{figure}

In an ordinary coaxial cable the electromagnetic wave propagates between the
metallic outer and inner conductors. The propagation velocity and the
impedance of the wave are determined by the electromagnetic inductance and 
capacitance, with $v_{0}=1/\sqrt{lc}$ and $Z=\sqrt{l/c}$. They are
close to the propagation velocity and impedance of electromagnetic
waves in free space. In an ordinary coaxial cable, the main part of the inductive energy is
stored in the magnetic field induced by the moving electrons. In contrast,
in nanotubes the kinetic energy of electrons exceeds their
magnetic energy and the inertia of the electrons slows down the
electromagnetic waves. The large kinetic 
inductance $l_{kin}$ is a consequence of the low
electron density of the nanotubes. 

The electromagnetic waves traveling along nanotubes are actually
electron density waves propagating in one dimension, that is, 1D plasmons. If
the MWNT is regarded as a transmission line, then its impedance will
determine the current-voltage characteristics for electron tunneling into
the tube over a mesoscopic tunnel junction, as described by the
environmental quantum fluctuations theory.\cite{Sonin,Devoret,IN} At low voltages, $I\propto
V^{\alpha +1}$, where $\alpha =2Re(Z)/R_{Q}$. At high voltages, the
tunnel junction capacitance $C_T$ starts to shunt the environmental
impedance $Z$, and the power law turns gradually into an Ohm's law
with a characteristic asymptotic region described by a $1/V$-type of
tail.\cite{TarkiainenPRL}

Figure \ref{TailLIN} illustrates the asymptotic approach of the 
$IV$-curve of sample AD1 towards Ohm's law. We  plot in Fig.~\ref{TailLIN} the 
deficit current $I_{d}=V/R_T-I$ which
is the deviation of the measured current $I$ from the Ohmic value given by $V/R_T$.
The solid line illustrates the fitting of the 
theoretical $1/V$-tail: $I_{d}={\frac{R_{Q}}{ Z} \left(
\frac{e}{2\pi C_{T}}\right) ^{2}{\frac{1 }{V}}}$,
which yields 
for this tube a typical value of $Z = 2.7$ k$\Omega $. 
Assuming  $C = 70$ aF/$\mu $m, in fact deduced from the Coulomb
blockade offset of the same fits,
we obtain $l \sim 0.5$ nH/$\mu$m. This inductance value is much higher than
expected for magnetic inductance ($l \sim 0.1$ pH/$\mu $m) but 
quite well in agreement with our
estimate for the kinetic inductance:\cite{TarkiainenPRL} 
$l_{kin} \sim 10$ nH/$\mu $m for
$M=4$. We thus conclude that, due to the large kinetic 
inductance, the propagation
velocity in MWNTs is about 1000 km/s and the impedance is 
on the order of a few k$\Omega $. 

\section{SUMMARY AND PROSPECTS}

Carbon nanotubes vary considerably in size and properties. Restricting
this variation to normal-sized AD and CVD multiwalled nanotubes, we
presented in the preceding sections our work on different device
configurations that can be realized from MWNTs. The difference in
structural quality between AD and CVD tubes was discussed, concluding
that AD tubes were preferable from the point of view of device
physics. We have
demonstrated the use of multiwalled carbon nanotubes as
nanoelectronics building blocks such as SETs, resistors and
transmission lines. Our SET made of two MWNTs demonstrated the
feasibility of more complex circuits. Under proper conditions, such as
separating the tube from the substrate, it is possible to minimize the noise
level and, in other aspects as well, to approach the theoretical limits of
performance. On the other hand, our present understanding of MWNTs as
transmission lines implies that their kinetic inductance severely
limits the propagation speed in them. Nevertheless, the great variety of
device structures that can be made from carbon nanotubes suggest that
they are among the most promising materials for future molecular electronics.

\section*{ACKNOWLEDGMENTS}

We acknowledge F. Hekking, E. Sonin, and A. Zaikin for interesting
discussions. This work was supported by the Academy of Finland and by the
Large Scale Installation Program ULTI-III of the European Union (HPRI-1999-CT-00050).

\end{document}